\documentclass[showpacs,prl,twocolumn,superscriptaddress]{revtex4}
\usepackage{graphicx,amsmath,color}
\begin{document}

\title{Collapse and revivals of the photon field in a
  Landau-Zener process}

\author{J. Keeling} 
\affiliation{Cavendish Laboratory, University of Cambridge, J J Thomson Ave., Cambridge, CB3 0HE, UK}
\author{V. Gurarie}
\affiliation{Department of Physics, University of Colorado, Boulder, CO 80309, USA}

\begin{abstract}
  We consider the evolution of a two-level system coupled to a photon
  field initially in a coherent state, as the energy of the two-level
  system is linearly varied through resonance with the photon field.
  At a fixed time after the resonance, the amplitude of the photon
  field is found to show a collapse and subsequent revivals as
  a function of rate of energy variation.
  Including decay of the photon field, we find that the observation of
  such collapse and revivals is near the technological limit of
  current cavity QED experiments but should be achievable.
\end{abstract}
\maketitle

The famous problem of Landau~\cite{landau32} and Zener~\cite{zener32}
concerns a two-level system whose parameters are varied so that an
anticrossing of energy levels occurs, and provides the probability
that the system will remain in an adiabatic state.
There has been much work since on finding generalisations of this
problem to many levels which can still be exactly solved,
e.g~\cite{brundobler93,shytov04,volkov04,volkov05,dobrescu06}.
Another generalisation of the Landau Zener process is to consider
multiple occupation of single particle levels, and then to find how
varying the single particle level energies affects the many particle
state\cite{altland07}.
These constitute a large class of possible problems, of which we study
a particular case relevant to cavity QED experiments%
~\cite{mabuchi02,blais04,brennecke07,auffeves03,raimond01}.
Cavity QED studies strong light-matter coupling between photons
confined in a cavity and matter (e.g. single atoms, or artificial
atoms such as quantum dots~\cite{khitrova07} or Josephson
junctions~\cite{blais04}).
We consider a cavity QED system in which the matter is driven through
a Landau-Zener transition, and find that this has dramatic
consequences for the state of the photon field: observing at a fixed
time after the anticrossing and changing the rate of energy variation,
there is a collapse and revival of the photon field amplitude.
These oscillations extend far into the regime of slow energy variation
--- the adiabatic limit --- in which the single particle Landau-Zener
probability shows no further change.

The scheme we propose consists of a two-level system coupled to a
photon mode in a cavity; the energy of the two-level system is varied
linearly in time, passing through resonance with the photon energy.
A related problem, but with a classical photon field, was considered
in Ref.~\cite{wubs05}; we also note that a generalisation of our model
to many two-level systems is analogous to models of interest
describing production of molecules in cold atomic gases, when varying
the molecular energy by a Feshbach resonance,
e.g.\cite{barankov05,altman05,altland07,sun07}.
If the initial cavity state were a number state then the result would
be the Landau-Zener energy level crossing:
The state with an excited two-level system and $n-1$ photons crosses
the state with an unexcited two-level system and $n$ photons;
the mutual repulsion of these states is given by $g \sqrt{n}$, where $g$ is
the radiation-matter coupling strength.
Varying the two-level system energy linearly in time as $E=\lambda t$
induces transitions between these states.
The standard Landau-Zener result~\cite{landau32,zener32} is that in terms of
the parameter $z = g^2/\lambda$, the probability of remaining in
the lowest energy state (rather than exciting the system) is $P=1-e^{-2 \pi z n }$.  
If the energy ramp is slow, or $z n \gg 1$, this quickly saturates at $1$.

\begin{figure}[htpb]
  \centering
  \includegraphics[width=0.9\columnwidth]{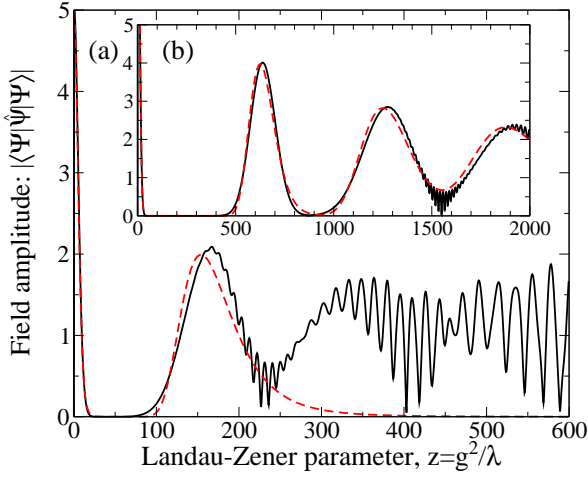}
  \caption{Dependence of the photon field intensity on the
    Landau-Zener parameter, $z=g^2/\lambda$.  Main figure, initial
    coherent state amplitude $\psi=5$; inset $\psi=10$.  The red
    dashed line is the approximation discussed in the text.}
  \label{fig:field}
\end{figure}

However, a natural state of the cavity is the coherent
state~\cite{scully97}, leading to a far more intricate behaviour.
The coherent state can be resolved onto a basis of number states, and
each number state evolves independently.
The effective radiation-matter coupling strength for $n$ photons is
$g\sqrt{n}$, and so the evolution of each photon number state is also
$n$ dependent.
In particular, each photon number state acquires a different phase
during the process; this phase difference continues to play an
important role even when the sweep velocity is so slow that $z\gg 1$ and
the probability for a single transition $P=1-e^{-2 \pi zn } \simeq 1$.
To measure the effect of these phase differences, the easiest quantity
to measure is the amplitude of the resultant photon field, $\langle
\hat{\psi} \rangle = \langle \Psi | \hat{\psi} | \Psi \rangle$.
How this is to be measured depends on the particular cavity QED system
used.
Some possible methods include Ramsey interferometry of a second atom
to probe the state of the cavity~\cite{raimond01}; homodyne
measurement of the photons leaking out of the cavity by interfering it
with a reference beam; and homodyne measurement inside the cavity by
detecting the interfering fields with a second atom~\cite{auffeves03}.
The dependence of this field amplitude on the rate of sweeping is shown
in Fig.~\ref{fig:field}, for two different values of the initial field
amplitude.

The photon field amplitude shows a collapse and subsequent revivals
arising from the phase difference between different number state
components of the coherent state.
The collapse occurs as, with slower driving, the phase difference
between different number states grows, and so their contributions to
the field amplitude interfere destructively.
At yet slower driving, the phase difference grows sufficiently that
subsequent number states are back in phase.
However, since the dependence of the phase on photon number is
nonlinear, these revivals are imperfect, and eventually cease.

Let us now discuss this behaviour more quantitatively; the Hamiltonian
of the problem is:
\begin{equation}
  \label{eq:1}
  \hat H = \omega_0 \, \hat \psi^\dagger \hat \psi + \frac{\lambda t}2  \hat \sigma_z + g \left( \hat \psi^\dagger  \hat \sigma^- + \hat \psi
    \, \hat \sigma^+ \right),
\end{equation}
At the beginning of the time evolution, the two-level system energy is
well below the photon energy and the system is initialised in a
coherent photon state,
\begin{math}
  \label{eq:2}
  \left| \Psi_i \right> 
  = 
  e^{-|\psi|^2/2}\sum_n (\psi^n/\sqrt{n!}) \left| n, \uparrow \right>.
\end{math}
Evolution under Eq.~(\ref{eq:1}) gives $ \left| n, \uparrow \right>
\to A_{n+1} \left| n, \uparrow \right> + B_{n+1} \left| n+1,
  \downarrow \right>$, where for evolution from $-T$ to $T$ for large
times $T$, the coefficients\cite{zener32,PhysRevA.53.4288} are:
\begin{align}
  \label{eq:lz-coeff}
  A_n &= e^{-n\pi z}, \\
  B_n &= 
   \frac{\sqrt{2\pi n z}e^{-n\pi z/2}}{\Gamma(1 -i n z)}
   e^{-i\left[
       \pi/4 + \lambda T^2/2 + 2zn \ln (\sqrt{\lambda} T)
     \right]}.
\end{align}
In the large $z$ limit where interesting behaviour is seen in
Fig.~\ref{fig:field}, $A_n \simeq 0$ and $|B_n|\simeq 1$ so we may
write $B_n = \exp(i\phi_n)$ with $\phi_n(T) = z n \left[ \ln(z n) -
  1\right] - 2 z n \ln(\sqrt{\lambda} T) + \phi_0$.
In this adiabatic limit, this phase is just the
integral of $E_n(t)$, the instantaneous energy of the eigenstate with
$n$ excitations.
The final many body state is then:
\begin{math}
  \label{eq:3}
  \left| \Psi \right> 
  = 
  e^{-|\psi|^2/2}\sum_n (\psi^n e^{i\phi_{n+1}(T)} / \sqrt{n!})
\left| n+1, \downarrow \right>.
\end{math}
The measurable field amplitude can then be written as
\begin{equation}
  \label{eq:4}
  \left< \hat\psi \right>
  =
  \psi e^{-|\psi|^2} \sum_n \frac{|\psi|^{2n}}{n!} \sqrt{\frac{n+2}{n+1}}
  e^{i(\phi_{n+2} - \phi_{n+1})}
\end{equation}

The existence and explanation of the collapse and revivals here is
related to the collapse and revival of Rabi
oscillations~\cite{narozhny81,gea-banacloche90,phoenix91}; this occurs
in a model like Eq.~(\ref{eq:1}) but without time-varying energies,
and collapse and revivals occur as a function of time.
Let us discuss the case $\left|\psi \right|^2 \gg 1$.
As in~\cite{narozhny81}, one can expand the phase difference
$\Delta \phi_n = \phi_{n+1} - \phi_n$ near $n=|\psi|^2$ (where the
amplitude of the terms in the sum peaks), giving
\begin{math}
  \Delta \phi_{|\psi|^2 + m} = \Delta \phi_{|\psi|^2} + z m / |\psi|^2
  - z m^2/(2|\psi|^4).
\end{math}
The revivals in Fig.~\ref{fig:field} occur when $\Delta \phi_n$ is the
same (modulo $2\pi$) for each term in Eq.~(\ref{eq:4}); for small $m$
this condition is $z = 2 \pi N |\psi|^2$, where $N$ is an integer
labelling the revival.
Near such revivals, after subtracting $2\pi N m$ from $\Delta
\phi_{|\psi|^2 +m}$, it can be written so it varies slowly with $m$:
\begin{displaymath}
  \Delta \phi_{|\psi|^2 + m}  -   2\pi N m
  = 
  \Delta \phi_{|\psi|^2} 
  + 
  \frac{(z-2\pi N |\psi|^2)m}{|\psi|^2} - \frac{zm^2}{2|\psi|^4}.
\end{displaymath}
This then allows the sum in Eq.~(\ref{eq:4}) to be replaced with an
integral over $m=n-|\psi|^2$.  Using the Gaussian approximation to the
Poisson distribution, yields:
\begin{displaymath}
  \left<  \hat\psi \right>
  =
  \int_{-\infty}^{\infty} 
  \frac{dm}{\sqrt{2\pi}}
  \exp\left[-\frac{m^2}{2|\psi|^2}
    + i \Delta \phi_{|\psi|^2 + m}
  \right].
\end{displaymath}
Evaluating this Gaussian integral, and summing over values of $N$ for
each revival gives the result:
\begin{equation}
  \label{eq:5}
  \left|\left<  \hat\psi \right>\right|
  =
  \frac{|\psi|}{\sqrt[4]{1 + z^2/|\psi|^4}}
  \!\sum_{N=0}^{N_{\text{max}}}\!\!
  \exp\left[
    \frac{- (z-2\pi N|\psi|^2)^2}{2|\psi|^2(1 + z^2/|\psi|^4)}
  \right].
\end{equation}
At very large $z$, the revivals disappear and the behaviour becomes
complex because terms of higher order in $m$ in the expansion of
$\Delta \phi_{|\psi|^2 + m}$ play a role when $z \gtrsim 3 |\psi|^3$.
This means revivals are not seen for $N>N_{\text{max}}  \approx |\psi|/2$.
Equation~(\ref{eq:5}) is shown in Fig.~\ref{fig:field} by the red
dashed line.

\begin{figure}[htpb]
  \centering
  \includegraphics[width=0.95\columnwidth]{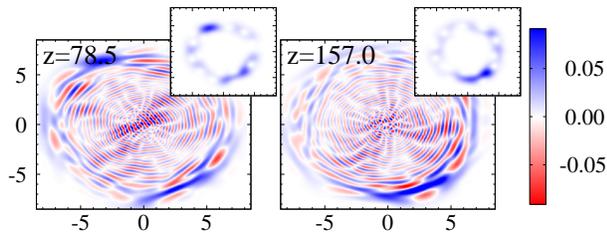}
  \caption{Wigner function of the final photon state for $\psi=5$ at values
    of $z$ corresponding to the first collapse and first revival.
    Inset: Glauber Q function on same scale for same parameters.
  }
  \label{fig:wigner}
\end{figure}

The revivals seen in the field amplitude do not indicate a complete
revival of the initial coherent state. 
This can be seen by considering the Wigner function~\cite{scully97},
$W(x,p) = \int \Psi^{\ast}(x+y) \Psi(x-y) e^{2ipy} dy/\pi$, where
$\Psi(x) = \langle x | \Psi \rangle$ is the position representation of
the photon wavefunction, and can be written in terms of Hermite
polynomials $H_n(x)$ (which correspond to number states) as:
\begin{equation}
  \label{eq:6}
    \Psi(x) 
    =
    \frac{e^{-(|\psi|^2+x^2)/2}}{\pi^{1/4}\psi}
    \sum_{n=1}^{\infty} 
    \frac{\psi^n e^{i\phi_n(T)}}{\sqrt{2^n} (n-1)!\sqrt{n}}
    H_n(x).
\end{equation}
As seen in Fig.~\ref{fig:wigner}, as soon as $z\gg 1$ the Wigner
function has many nodes, and describes a highly non-classical state,
unlike the initial coherent state whose Wigner function is a
Gaussian~\cite{scully97}.
We note that unlike the collapse and revival of Rabi
oscillations~\cite{gea-banacloche90,phoenix91} this non-classical
state is not associated with entanglement between the two-level system
and photon field~\cite{brune96} at the end of the dynamics, although
transient entanglement does occur during the sweep.
Since the collapse and revival occur in the adiabatic limit the
two-level system always ends in a pure state and so (in the absence of
decay) the photon state is a pure but non-classical state.
The complete Wigner function can be measured by quantum state
tomography~\cite{vogel89,smithey93}; such measurements have recently
been performed in cavity QED experiments~\cite{bertet02,houck07}.

In a real cavity QED experiment there is decoherence due to escape of
photons out of the cavity, and decay of the two-level system without
emitting a photon; these may present an obstacle to observing the
revivals, and so we next investigate the maximum photon loss rate
$\kappa$ and non-radiative decay rate $\gamma$ that can be tolerated.
%
With a non-zero decay rate $\kappa$, one must consider an finite
duration of the level crossing, and to maximise the signal one should
make the duration as short as possible consistent with accumulating
the necessary phase.
A naive estimate of the effect of decay comes from $(\kappa
t^{\ast},\gamma t^{\ast})$ where $t^{\ast}$ is the time for the level
crossing to occur, given by $\lambda t^{\ast}
\simeq g |\psi|$, and $g |\psi|$ is the characteristic
coupling strength for the coherent photon state.
Decay is weak if $(\kappa t^{\ast}, \gamma
t^{\ast}) \ll 1$, which gives $(\kappa,\gamma)/g \ll (\lambda/g^2) /
|\psi|$.
For this condition to be satisfied at
the first revival, when $g^2/\lambda = z = 2 \pi |\psi|^2$, one
requires that $(\kappa,\gamma)/g \ll 1 / (2 \pi |\psi|^3)$.
Away from the level crossing the effects of decay are less serious;
decay after the crossing attenuates the amplitude of the
signal, decay beforehand reduces the amplitude of the
initial photon field amplitude.
This naive estimate is sufficient for the effects of non-radiative
decay, but not for loss of photons.

\begin{figure}[htpb]
  \centering
  \includegraphics[width=1.0\columnwidth]{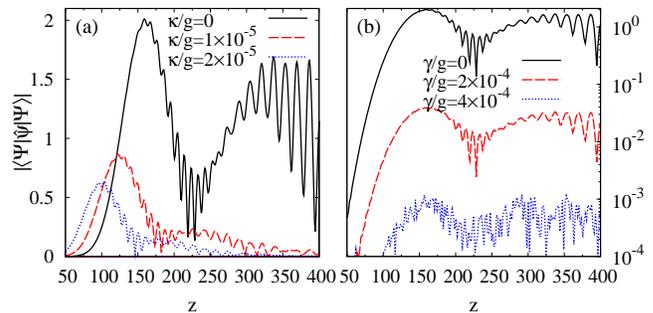}
  \caption{Effect of decay on $z$ dependence of field amplitude.  Left
    panel: effect of photon leakage, $\kappa$.  Right panel: effect of
    non-radiative decay $\gamma$ (with field amplitude plotted on
    logarithmic scale).  Plotted for $\psi=5$ and a time range of
    $-10^4 < g t < 10^4$. }
  \label{fig:kappa}
\end{figure}

To investigate the effects of decay quantitatively, we solve
numerically the density matrix equation of motion:
\begin{equation}
  \label{eq:7}
  \partial_t \hat{\rho} = -i \left[ \hat{H}, \hat{\rho} \right]
  +
  \mathcal{L}_\kappa[\hat\rho]
  +
  \mathcal{L}_\gamma[\hat\rho]
\end{equation}
where
\begin{math}
  \mathcal{L}_\kappa[\hat\rho]
  =
  (\kappa/2)
  [
  \hat\psi^{\dagger} \hat\psi \hat{\rho}
  +
  \hat{\rho} \hat\psi^{\dagger} \hat\psi
  -
  2 \hat\psi \hat{\rho} \hat\psi^{\dagger} 
  ]
\end{math}
describes loss of photons and
\begin{math}
  \mathcal{L}_\gamma[\hat\rho]
  =
  (\gamma/2)
  [
  \hat\sigma_+ \hat\sigma_0 \hat{\rho}
  +
  \hat{\rho} \hat\sigma_+ \hat\sigma_-
  -
  2 \hat\sigma_- \hat{\rho} \hat\sigma_+ 
  ]
\end{math}
describes non-radiative decay of spins~\cite{scully97}.
The numerical solution of this equation, shown in
Fig.~\ref{fig:kappa}, reveals a greater effect of photon loss than the
naive estimate above.
For $\psi=5$, the maximum permissible decay rates is $\kappa/g \simeq
2 \times 10^{-5}$, $\gamma/g \simeq 2\times 10^{-4}$, compared to the
naive estimate $1 / (2 \pi |\psi|^3) \simeq 10^{-3}$.

To understand this enhancement of decay, it is convenient to calculate
the effect of photon decay in the adiabatic limit, such that the state
$\left| n-1, \uparrow \right>$ evolves to:
\begin{align}
  \label{eq:8}
     \left|n, +\right> &=  \left[
    \cos(\theta_n)
    \left|n, \downarrow\right>
    +
    \sin(\theta_n)
    \left|n-1, \uparrow\right>
  \right]
  \\
   \label{eq:defn-coeff}
  \cos(\theta_n) &= 
  \sqrt{\frac{1}{2}\left[
      1 + \frac{\tau}{\sqrt{ n +\tau^2}}\right]},
  \quad
  \tau = \frac{\lambda t}{2g}.
\end{align}
The adiabatic approximation is appropriate even with photon
decay because the probability that loss of a photon swaps between
the $|+\rangle$ and $|-\rangle$ subspaces is small.
This probability, of switching from $|n, +\rangle$ to $|n-1, -\rangle$
is given by:
\begin{math}
  P_{-+} =
  |\langle n-1,-| \hat \psi | n, + \rangle|^2 / 
  \langle n,+| \hat \psi^{\dagger} \hat \psi | n, + \rangle
\end{math}
and can be written in terms of the $\theta_n, \tau$ of
Eq.~(\ref{eq:defn-coeff}) as:
\begin{align}
  \label{eq:9}
  P_{-+} &=
  \frac{|\cos(\theta_n) \sin(\theta_{n-1}) \sqrt{n} -
    \sin(\theta_n) \cos(\theta_{n-1}) \sqrt{n-1}|^2}{n - \sin^2(\theta_n)}
  \nonumber\\
  &\approx
  \frac{1}{16 n^2} \left[ 
    \frac{\tau \sqrt{n}}{n+\tau^2} - \sqrt{\frac{n}{n+\tau^2}}
  \right]^2
  \leq
  \frac{1}{16 n^2} \frac{27}{16}.
\end{align}
where the last expression makes use of approximations for $n \gg 1$.
It is thus clear that for large $n$, the probability of leaving
the adiabatic subspace is small.

In this adiabatic subspace, the density matrix equation gives a closed
set of equations for off-diagonal terms, $\Lambda_n = \left<
  n-1,+ | \hat{\rho} | n,+ \right>$, and the measured field amplitude
is $\langle \Psi | \hat \psi | \Psi \rangle = \sum_n \sqrt{n}
\Lambda_n$.
The effect of the Hamiltonian is  phase evolution
of each $\Lambda_n$, with the final phase gain of $\Lambda_n$ being
$\Delta \phi_{n-1}(T)$ as in Eq.~(\ref{eq:4}).
Including also the matrix elements due to photon decay, one can
write:
\begin{equation}
  \label{eq:10}
  \frac{d \Lambda_n}{dt} = i \frac{d \Delta \phi_{n-1}}{dt}  \Lambda_n
  - \kappa \left[ \left(n - \frac{1}{2}\right) \Lambda_n 
    - \sqrt{n^2+n} \Lambda_{n+1} \right]
\end{equation}
For the range of parameters shown in Fig.~\ref{fig:kappa}(a), the
results of numerical evaluation of this equation and of the full
problem, Eq.~(\ref{eq:7}) cannot be distinguished by eye.

If $\kappa=0$, the solution for $\Lambda_n$ recovers
Eq.~(\ref{eq:4}).
Alternatively, if $\Delta \phi_n$ is time independent, then for an
initial coherent state, the time evolution can also be
exactly solved~\cite{scully97} and the field amplitude decays as
$\psi(t) = \psi(0) e^{-\kappa t/2}$.
It may appear surprising that the characteristic decay rate
is $\kappa$, while the apparent rate in
Eq.~(\ref{eq:10}) is $\kappa n$; the explanation is that the order $n$
contributions from the $\Lambda_n$ and $\Lambda_{n+1}$ terms cancel.
However, with a time-dependent $\Delta \phi_n$, the $\Lambda_n$ and
$\Lambda_{n+1}$ terms pick up a relative phase difference, and the
cancellation of the order $n$ term fails.
Thus the characteristic decay rate is $\kappa n \simeq \kappa
|\psi|^2$ during the level-crossing time, hence the actual requirement
for small decay is $\kappa/g \ll 1 / (2 \pi |\psi|^5)$; for $\psi=5$
this gives $\kappa/g \ll 5 \times 10^{-5}$ (cf Fig.~\ref{fig:kappa}).

Of the various realisations of cavity QED, those with parameters
closest to those required by the constraints on $\kappa/g, \gamma/g$
are either Josephson junctions coupled to stripline
resonators\cite{blais04}, or Rydberg atoms in microwave
cavities~\cite{kuhr07}.  
For Rydberg atoms the values of $\kappa/g$, $\gamma/g$ currently
achievable are very close to those required, but the limiting factor
is the short time taken for an atom to pass through the cavity.

Finally, let us present a more quantitative understanding of the
``enhanced decay'' by comparing the perturbative solution of
Eq.~(\ref{eq:10}) to the naive expectation:
\begin{equation}
  \label{eq:11}
  \left< \Psi|\hat\psi(\kappa,\psi_0)|\Psi \right>_{\text{naive}}
  =
  \left< \Psi|\hat\psi(0,\psi_0 e^{-\kappa T/2})|\Psi \right>
  e^{-\kappa T/2}
\end{equation}
This  naive decay describes decay of the initial
coherent state before the anticrossing (thus shifting the revivals to
smaller values of $z$), and decay of the field amplitude after the
anticrossing.
The perturbative solution (to leading order in $\kappa$) of
Eq.~(\ref{eq:10}) can be easily found by writing $\Lambda_n
=\tilde{\Lambda}_n\exp[-i \Delta \phi_{n-1}(t)]$, and then ignoring
the $\kappa$ dependence of $\tilde{\Lambda}_n$ on the right hand side
of Eq.~(\ref{eq:10}).
It will be convenient to write the initial conditions for $\Lambda_n$
as $\Lambda_n = P_n \psi / \sqrt{n}$, where $P_n = e^{-|\psi|^2}
|\psi|^{2(n-1)}/(n-1)!$ is the probability of having $n-1$ photons.
Writing
\begin{math}
  \delta \langle \Psi|\hat\psi|\Psi \rangle = \langle \Psi|\hat\psi|\Psi
\rangle_{\text{naive}} - \langle \Psi|\hat\psi|\Psi \rangle
\end{math}
and expanding both Eq.~(\ref{eq:11}) and the solution to Eq.~(\ref{eq:10})
to leading order in $\kappa$, one has:
\begin{multline}
  \delta \langle \Psi|\hat\psi|\Psi \rangle
  = 
   \kappa |\psi|^2 
  \sum_n  P_n \psi \times \\
   \left[\vphantom{\int}
    T e^{i[\Delta\phi_{n-1}(-T) - \Delta\phi_{n-1}(T)]}  + T e^{i[\Delta\phi_{n}(-T) - \Delta\phi_{n}(T)]}
     \right.\\\left. -
    e^{i[\Delta\phi_{n}(-T) - \Delta\phi_{n-1}(T)]}
    \int_{-T}^T   e^{i[\Delta\phi_{n-1} - \Delta\phi_{n}]} dt
    \right]
\end{multline}
Assuming $n \gg 1$, one can simplify the expressions for the
phase differences appearing here to give:
\begin{multline}
  \label{eq:perturbative-final}
  \delta \langle \Psi|\hat\psi|\Psi \rangle
  = 
  \kappa |\psi|^2
  \sum_n P_n \psi 
  e^{-i z \ln(T^2/zn) + iz/2n}\times
  \\
  \int_{-T}^T
  dt \left[ 
    \cos\left(\frac{z}{2n}\right)
    -
    \cos\left(\frac{z}{2n} 
      \frac{\sqrt{\lambda} t/2}{\sqrt{z n + \lambda
          t^2/4}}
    \right)
  \right]
\end{multline}
Finally, note that the integral in Eq.~(\ref{eq:perturbative-final})
has a finite limit as $T \to \infty$; this means that the naive decay
that was subtracted fully describes the decay at long times, and
Eq.~(\ref{eq:perturbative-final}) describes only the extra
contribution that occurs during the crossing.
Taking $T\to\infty$, the integral can be found in terms of a
Bessel function yielding:
\begin{multline}
  \label{eq:perturbative-final-dimless}
  \delta\!\langle \Psi|\hat\psi|\Psi \rangle
  = 
  -
   \kappa |\psi|^2 \pi \times
  \\
  \sum_n P_n \psi
  e^{-i z \ln\left(\frac{T^2}{zn}\right)
    + i \left( \frac{z}{2n} \right)}
  \sqrt{\frac{z^3}{n\lambda}} 
   J_1\left(\frac{z}{2n}\right)
\end{multline}
Using the asymptotic form for the Bessel function, this expression
shows that the characteristic scale of the extra decay is given by:
\begin{math}
  \delta\!\langle \Psi|\hat\psi|\Psi \rangle 
  \propto
  \kappa |\psi|^2
  \psi z / \sqrt{\lambda}
  \propto 
  (\kappa/g)
  |\psi|^3 z^{3/2}
\end{math}.
This extra term explains both the scale of the extra decay observed in
Fig.~\ref{fig:kappa}(a), and also the $z$ dependence of the decay
visible in that figure.

In summary we have proposed an experiment in which collapse and
revivals of a coherent field amplitude can be seen as a function of
varying sweep rate in a Landau-Zener level crossing problem.
Considering a leaky cavity and non-radiative decay, this effect
continues to survive as long as decay is weak enough, however the
sense of ``weak enough'', $\kappa/g \le 10^{-5}$, differs from the
naive expectation.
As such, this proposed experiment is at the limit of the capability of
current experiments, and so provides a dramatic consequence to be
observed by any system far in the strong-coupling limit

\begin{acknowledgments}
 We would like to thank B. D. Simons, K.
Lehnert and S. Gleyzes for helpful discussions. J.K. acknowledges
financial support from Pembroke College Cambridge. V.G. acknowledges
support by NSF via grant DMR-0449521. 
\end{acknowledgments}


\begin{thebibliography}{29}
\expandafter\ifx\csname natexlab\endcsname\relax\def\natexlab#1{#1}\fi
\expandafter\ifx\csname bibnamefont\endcsname\relax
  \def\bibnamefont#1{#1}\fi
\expandafter\ifx\csname bibfnamefont\endcsname\relax
  \def\bibfnamefont#1{#1}\fi
\expandafter\ifx\csname citenamefont\endcsname\relax
  \def\citenamefont#1{#1}\fi
\expandafter\ifx\csname url\endcsname\relax
  \def\url#1{\texttt{#1}}\fi
\expandafter\ifx\csname urlprefix\endcsname\relax\def\urlprefix{URL }\fi
\providecommand{\bibinfo}[2]{#2}
\providecommand{\eprint}[2][]{\url{#2}}

\bibitem[{\citenamefont{Landau}(1932)}]{landau32}
\bibinfo{author}{\bibfnamefont{L.~D.} \bibnamefont{Landau}},
  \bibinfo{journal}{Physik. Z. Sowjetunion} \textbf{\bibinfo{volume}{2}},
  \bibinfo{pages}{46} (\bibinfo{year}{1932}).

\bibitem[{\citenamefont{Zener}(1932)}]{zener32}
\bibinfo{author}{\bibfnamefont{C.}~\bibnamefont{Zener}},
  \bibinfo{journal}{Proc. Roy. Soc. A} \textbf{\bibinfo{volume}{137}},
  \bibinfo{pages}{696} (\bibinfo{year}{1932}).

\bibitem[{\citenamefont{Brundobler and Elser}(1993)}]{brundobler93}
\bibinfo{author}{\bibfnamefont{S.}~\bibnamefont{Brundobler}} \bibnamefont{and}
  \bibinfo{author}{\bibfnamefont{V.}~\bibnamefont{Elser}}, \bibinfo{journal}{J.
  Phys. A:Math. Gen.} \textbf{\bibinfo{volume}{26}}, \bibinfo{pages}{1211}
  (\bibinfo{year}{1993}).

\bibitem[{\citenamefont{Shytov}(2004)}]{shytov04}
\bibinfo{author}{\bibfnamefont{A.~V.} \bibnamefont{Shytov}},
  \bibinfo{journal}{Phys. Rev. A} \textbf{\bibinfo{volume}{70}},
  \bibinfo{pages}{052708} (\bibinfo{year}{2004}).

\bibitem[{\citenamefont{Volkov and Ostrovsky}(2004)}]{volkov04}
\bibinfo{author}{\bibfnamefont{M.~V.} \bibnamefont{Volkov}} \bibnamefont{and}
  \bibinfo{author}{\bibfnamefont{V.~N.} \bibnamefont{Ostrovsky}},
  \bibinfo{journal}{J. Phys. B: At. Mol. Opt. Phys.}
  \textbf{\bibinfo{volume}{37}}, \bibinfo{pages}{4069} (\bibinfo{year}{2004}).

\bibitem[{\citenamefont{Volkov and Ostrovsky}(2005)}]{volkov05}
\bibinfo{author}{\bibfnamefont{M.~V.} \bibnamefont{Volkov}} \bibnamefont{and}
  \bibinfo{author}{\bibfnamefont{V.~N.} \bibnamefont{Ostrovsky}},
  \bibinfo{journal}{J. Phys. B: At. Mol. Opt. Phys.}
  \textbf{\bibinfo{volume}{38}}, \bibinfo{pages}{907} (\bibinfo{year}{2005}).

\bibitem[{\citenamefont{Dobrescu and Sinitsyn}(2006)}]{dobrescu06}
\bibinfo{author}{\bibfnamefont{B.~E.} \bibnamefont{Dobrescu}} \bibnamefont{and}
  \bibinfo{author}{\bibfnamefont{N.~A.} \bibnamefont{Sinitsyn}},
  \bibinfo{journal}{J. Phys. B: At. Mol. Opt. Phys.}
  \textbf{\bibinfo{volume}{39}}, \bibinfo{pages}{1253} (\bibinfo{year}{2006}).

\bibitem[{\citenamefont{Altland and Gurarie}(2008)}]{altland07}
\bibinfo{author}{\bibfnamefont{A.}~\bibnamefont{Altland}} \bibnamefont{and}
  \bibinfo{author}{\bibfnamefont{V.}~\bibnamefont{Gurarie}},
  \bibinfo{journal}{Phys. Rev. Lett.} \textbf{\bibinfo{volume}{100}},
  \bibinfo{pages}{063602} (\bibinfo{year}{2008}).

\bibitem[{\citenamefont{Mabuchi and Doherty}(2002)}]{mabuchi02}
\bibinfo{author}{\bibfnamefont{H.}~\bibnamefont{Mabuchi}} \bibnamefont{and}
  \bibinfo{author}{\bibfnamefont{A.~C.} \bibnamefont{Doherty}},
  \bibinfo{journal}{Science} \textbf{\bibinfo{volume}{298}},
  \bibinfo{pages}{1372} (\bibinfo{year}{2002}).

\bibitem[{\citenamefont{Blais et~al.}(2004)\citenamefont{Blais, Huang,
  Wallraff, Girvin, and Schoelkopf}}]{blais04}
\bibinfo{author}{\bibfnamefont{A.}~\bibnamefont{Blais}}, \bibnamefont{et~al.},
  \bibinfo{journal}{Phys. Rev. A} \textbf{\bibinfo{volume}{69}},
  \bibinfo{pages}{062320} (\bibinfo{year}{2004}).

\bibitem[{\citenamefont{Brennecke et~al.}(2007)\citenamefont{Brennecke, Donner,
  Ritter, Bourdel, K\"ohl, and Esslinger}}]{brennecke07}
\bibinfo{author}{\bibfnamefont{F.}~\bibnamefont{Brennecke}},
  \bibnamefont{et~al.}, \bibinfo{journal}{Nature}
  \textbf{\bibinfo{volume}{450}}, \bibinfo{pages}{268} (\bibinfo{year}{2007}).

\bibitem[{\citenamefont{Auffeves et~al.}(2003)\citenamefont{Auffeves, Maioli,
  Meunier, Gleyzes, Nogues, Brune, Raimond, and Haroche}}]{auffeves03}
\bibinfo{author}{\bibfnamefont{A.}~\bibnamefont{Auffeves}},
  \bibnamefont{et~al.}, \bibinfo{journal}{Phys. Rev. Lett.}
  \textbf{\bibinfo{volume}{91}}, \bibinfo{pages}{230405}
  (\bibinfo{year}{2003}).

\bibitem[{\citenamefont{Raimond et~al.}(2001)\citenamefont{Raimond, Brune, and
  Haroche}}]{raimond01}
\bibinfo{author}{\bibfnamefont{J.~M.} \bibnamefont{Raimond}},
  \bibinfo{author}{\bibfnamefont{M.}~\bibnamefont{Brune}}, \bibnamefont{and}
  \bibinfo{author}{\bibfnamefont{S.}~\bibnamefont{Haroche}},
  \bibinfo{journal}{Rev. Mod. Phys.} \textbf{\bibinfo{volume}{73}},
  \bibinfo{eid}{565} (\bibinfo{year}{2001}).

\bibitem[{\citenamefont{Khitrova et~al.}(2007)\citenamefont{Khitrova, Gibbs,
  Kira, Koch, and Scherer}}]{khitrova07}
\bibinfo{author}{\bibfnamefont{G.}~\bibnamefont{Khitrova}},
  \bibnamefont{et~al.}, \bibinfo{journal}{Nature Physics}
  \textbf{\bibinfo{volume}{3}} (\bibinfo{year}{2007}).

\bibitem[{\citenamefont{Wubs et~al.}(2005)\citenamefont{Wubs, Saito, Kohler,
  Kayanuma, and H\"anggi}}]{wubs05}
\bibinfo{author}{\bibfnamefont{M.}~\bibnamefont{Wubs}}, \bibnamefont{et~al.},
  \bibinfo{journal}{New Journal of Physics} \textbf{\bibinfo{volume}{7}},
  \bibinfo{pages}{218} (\bibinfo{year}{2005}).

\bibitem[{\citenamefont{Barankov and Levitov}()}]{barankov05}
\bibinfo{author}{\bibfnamefont{R.~A.} \bibnamefont{Barankov}} \bibnamefont{and}
  \bibinfo{author}{\bibfnamefont{L.~S.} \bibnamefont{Levitov}},
  \eprint{cond-mat/0506323}.

\bibitem[{\citenamefont{Altman and Vishwanath}(2005)}]{altman05}
\bibinfo{author}{\bibfnamefont{E.}~\bibnamefont{Altman}} \bibnamefont{and}
  \bibinfo{author}{\bibfnamefont{A.}~\bibnamefont{Vishwanath}},
  \bibinfo{journal}{Phys. Rev. Lett.} \textbf{\bibinfo{volume}{95}},
  \bibinfo{pages}{110404} (\bibinfo{year}{2005}).

\bibitem[{\citenamefont{Sun et~al.}()\citenamefont{Sun, Abanov, and
  Pokrovsky}}]{sun07}
\bibinfo{author}{\bibfnamefont{D.}~\bibnamefont{Sun}},
  \bibinfo{author}{\bibfnamefont{A.}~\bibnamefont{Abanov}}, \bibnamefont{and}
  \bibinfo{author}{\bibfnamefont{V.~L.} \bibnamefont{Pokrovsky}},
  \bibinfo{note}{arXiv:0707.3630}.

\bibitem[{\citenamefont{Scully and Zubairy}(1997)}]{scully97}
\bibinfo{author}{\bibfnamefont{M.~O.} \bibnamefont{Scully}} \bibnamefont{and}
  \bibinfo{author}{\bibfnamefont{M.~S.} \bibnamefont{Zubairy}},
  \emph{\bibinfo{title}{Quantum Optics}} (\bibinfo{publisher}{Cambridge
  University Press}, \bibinfo{address}{Cambridge}, \bibinfo{year}{1997}).

\bibitem[{\citenamefont{Vitanov and Garraway}(1996)}]{PhysRevA.53.4288}
\bibinfo{author}{\bibfnamefont{N.~V.} \bibnamefont{Vitanov}} \bibnamefont{and}
  \bibinfo{author}{\bibfnamefont{B.~M.} \bibnamefont{Garraway}},
  \bibinfo{journal}{Phys. Rev. A} \textbf{\bibinfo{volume}{53}},
  \bibinfo{pages}{4288} (\bibinfo{year}{1996}).

\bibitem[{\citenamefont{Narozhny et~al.}(1981)\citenamefont{Narozhny,
  Sanchez-Mondragon, and Eberly}}]{narozhny81}
\bibinfo{author}{\bibfnamefont{N.~B.} \bibnamefont{Narozhny}},
  \bibinfo{author}{\bibfnamefont{J.~J.} \bibnamefont{Sanchez-Mondragon}},
  \bibnamefont{and} \bibinfo{author}{\bibfnamefont{J.~H.}
  \bibnamefont{Eberly}}, \bibinfo{journal}{Phys. Rev. A}
  \textbf{\bibinfo{volume}{23}}, \bibinfo{pages}{236} (\bibinfo{year}{1981}).

\bibitem[{\citenamefont{Gea-Banacloche}(1990)}]{gea-banacloche90}
\bibinfo{author}{\bibfnamefont{J.}~\bibnamefont{Gea-Banacloche}},
  \bibinfo{journal}{Phys. Rev. Lett.} \textbf{\bibinfo{volume}{65}},
  \bibinfo{pages}{3385} (\bibinfo{year}{1990}).

\bibitem[{\citenamefont{Phoenix and Knight}(1991)}]{phoenix91}
\bibinfo{author}{\bibfnamefont{S.~J.~D.} \bibnamefont{Phoenix}}
  \bibnamefont{and} \bibinfo{author}{\bibfnamefont{P.~L.}
  \bibnamefont{Knight}}, \bibinfo{journal}{Phys. Rev. Lett.}
  \textbf{\bibinfo{volume}{66}}, \bibinfo{pages}{2833} (\bibinfo{year}{1991}).

\bibitem[{\citenamefont{Brune et~al.}(1996)\citenamefont{Brune, Hagley, Dreyer,
  Ma\^itre, Maali, Wunderlich, Raimond, and Haroche}}]{brune96}
\bibinfo{author}{\bibfnamefont{M.}~\bibnamefont{Brune}}, \bibnamefont{et~al.},
  \bibinfo{journal}{Phys. Rev. Lett.} \textbf{\bibinfo{volume}{77}},
  \bibinfo{pages}{4887} (\bibinfo{year}{1996}).

\bibitem[{\citenamefont{Vogel and Risken}(1989)}]{vogel89}
\bibinfo{author}{\bibfnamefont{K.}~\bibnamefont{Vogel}} \bibnamefont{and}
  \bibinfo{author}{\bibfnamefont{H.}~\bibnamefont{Risken}},
  \bibinfo{journal}{Phys. Rev. A} \textbf{\bibinfo{volume}{40}},
  \bibinfo{pages}{2847} (\bibinfo{year}{1989}).

\bibitem[{\citenamefont{Smithey et~al.}(1993)\citenamefont{Smithey, Beck,
  Raymer, and Faridani}}]{smithey93}
\bibinfo{author}{\bibfnamefont{D.~T.} \bibnamefont{Smithey}},
  \bibinfo{author}{\bibfnamefont{M.}~\bibnamefont{Beck}},
  \bibinfo{author}{\bibfnamefont{M.~G.} \bibnamefont{Raymer}},
  \bibnamefont{and} \bibinfo{author}{\bibfnamefont{A.}~\bibnamefont{Faridani}},
  \bibinfo{journal}{Phys. Rev. Lett.} \textbf{\bibinfo{volume}{70}},
  \bibinfo{pages}{1244} (\bibinfo{year}{1993}).

\bibitem[{\citenamefont{Bertet et~al.}(2002)\citenamefont{Bertet, Auffeves,
  Maioli, Osnaghi, Meunier, Brune, Raimond, and Haroche}}]{bertet02}
\bibinfo{author}{\bibfnamefont{P.}~\bibnamefont{Bertet}}, \bibnamefont{et~al.},
  \bibinfo{journal}{Phys. Rev. Lett.} \textbf{\bibinfo{volume}{89}},
  \bibinfo{pages}{200402} (\bibinfo{year}{2002}).

\bibitem[{\citenamefont{Houck et~al.}(2007)\citenamefont{Houck, Schuster,
  Gambetta, Schreier, Johnson, Chow, Frunzio, Majer, Devoret, Girvin
  et~al.}}]{houck07}
\bibinfo{author}{\bibfnamefont{A.~A.} \bibnamefont{Houck}},
  \bibnamefont{et~al.}, \bibinfo{journal}{Nature}
  \textbf{\bibinfo{volume}{449}}, \bibinfo{pages}{328} (\bibinfo{year}{2007}).

\bibitem[{\citenamefont{Kuhr et~al.}(2007)\citenamefont{Kuhr, Gleyzes, Guerlin,
  Bernu, Hoff, Del\'{e}glise, Osnaghi, Brune, Raimond, Haroche
  et~al.}}]{kuhr07}
\bibinfo{author}{\bibfnamefont{S.}~\bibnamefont{Kuhr}}, \bibnamefont{et~al.},
  \bibinfo{journal}{Appl. Phys. Lett.} \textbf{\bibinfo{volume}{90}},
  \bibinfo{eid}{164101} (\bibinfo{year}{2007}).

\end{thebibliography}

\end{document}